\documentclass[letterpaper,12pt]{article}   
\usepackage{amsmath,amssymb}
\usepackage{bm}
\usepackage[latin1]{inputenc}
\usepackage{dcolumn}
\usepackage{graphicx}
\newcommand{\matrise}[1]{\begin{bmatrix} #1 \end{bmatrix}}
\newcommand{\diff}[0]{\text{d}}
\usepackage{osajnl2} 
\usepackage[draft]{hyperref} 

\begin{document}

\title{Inverse scattering of dispersive stratified structures}

\author{Johannes Skaar,$^{1,2,*}$ and Magnus W. Haakestad,$^3$ }
\address{$^1$Department of Electronics and Telecommunications, Norwegian University of Science and Technology, NO-7491 Trondheim, Norway}
\address{$^2$University Graduate Center, NO-2027 Kjeller, Norway}
\address{$^3$Norwegian Defence Research Establishment (FFI), P O Box 25, NO-2027 Kjeller, Norway}
\address{$^*$Corresponding author: johannes.skaar@ntnu.no}


\begin{abstract}
We consider the inverse scattering problem of retrieving the structural parameters of a stratified medium consisting of dispersive materials, given knowledge of the complex reflection coefficient in a finite frequency range. It is shown that the inverse scattering problem does not have a unique solution in general. When the dispersion is sufficiently small, such that the time-domain Fresnel reflections have durations less than the round-trip time in the layers, the solution is unique and can be found by layer peeling. Numerical examples with dispersive and lossy media are given, demonstrating the usefulness of the method for e.g. THz technology.
\end{abstract}


\maketitle 

\section{Introduction}
The ability to detect and identify materials, hidden behind barriers, has potential applications within security and non-destructive testing \cite{kemp:2003,moller:2009}. The THz range of the electromagnetic spectrum is particularly attractive for these applications, because many barrier materials, such as clothing, plastic, and paper only attenuate THz waves moderately, while other materials,
such as explosives and related compounds have characteristic spectroscopic fingerprints in the THz region \cite{kemp:2003, shen:2005}. A relevant geometry for these applications is the reflection geometry in which a pulsed or CW signal is sent towards an unknown structure and the amplitude and phase of the reflected signal is detected \cite{van_Rheenen:2011}. The task is then to deduce the structure from the measured reflection coefficient. The situation is simplified in the effectively one-dimensional case, where the electromagnetic properties of the structure only vary in one direction. Retrieval of the structure parameters then becomes a one-dimensional inverse-scattering problem. However, because the relevant materials are lossy in the THz range, they are also dispersive, according to the Kramers--Kronig relations. Thus when applying inverse-scattering algorithms to the THz range one must take into account absorption and material dispersion.

There are several formulations and algorithms for the one-dimensional inverse scattering problem \cite{GLM, marchenko, brucksteintrans, yagle_levy, songfbg, frangosJ91, feced, skaarlp, rosenthal, skaar_waagaard, waagaardmulti}. In particular, the layer peeling algorithms have turned out to be very efficient, and used in a wide range of applications \cite{brucksteintrans, yagle_levy, feced, skaarlp, rosenthal, skaar_waagaard, waagaardmulti, skaar_erdogan}. These algorithms are based on the following, simple fact: Consider the time-domain reflection impulse response of a layered structure. By causality, the leading edge is only dependent on the first layer, since the wave has not had time to feel the presence of the other layers. Thus one can identify the first layer of the structure from the impulse response. This information can be used to remove the influence from the layer, which leads to synthetic reflection data with the first layer peeled off. This procedure can be continued until the complete structure 
has been identified.

In this work we generalize the layer-peeling algorithm to dispersive stratified structures. Provided the material dispersion is sufficiently small, such that the time-domain Fresnel reflections have durations less than the round-trip time in each layer, we can uniquely reconstruct the refractive indices of the structure (Section II). The method is illustrated by numerical examples in Section III. Finally, in Section IV we prove that for larger dispersion, the inverse scattering problem does not have a unique solution in general. This is because one cannot distinguish between the non-instantaneous temporal response of the medium itself (due to dispersion), and the temporal response due to the stratification (caused by reflections at the layer boundaries). Thus, extra information is needed, such as an upper bound for the dispersion combined with a lower bound for the layer thicknesses.

\section{Transfer matrix model and layer peeling}\label{layerpeelingdisp}
\begin{figure}
\centerline{\includegraphics[width=10cm]{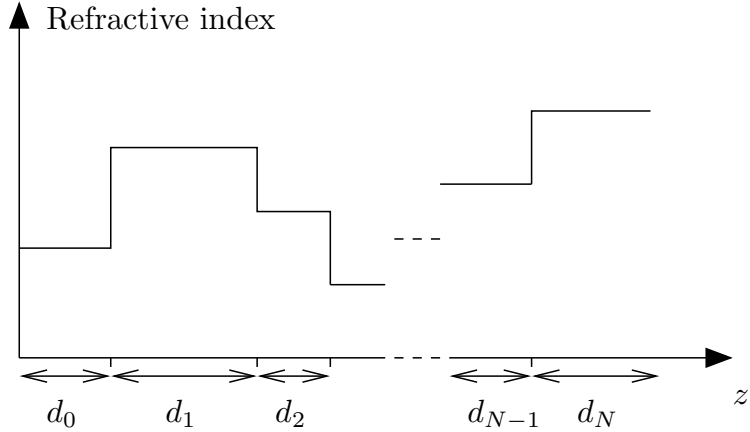}}
\caption{A planar structure consisting of $N+1$ layers. The thicknesses and refractive indices of the layers are $d_j$ and $n_j(\omega)$, respectively.} 
\label{fig:layers}
\end{figure}

We first describe the model of the stratified medium. Consider a layered, planar structure, consisting of $N+1$ layers with refractive indices $n_j(\omega)$ and thicknesses $d_j$, see Fig. \ref{fig:layers}. Here the index $j=0, 1, \ldots, N$ labels the layer. The light propagation in this structure is conveniently modeled using transfer matrices. For simplicity we limit the analysis to normal incidence. The transfer matrix of the transition from refractive index $n_{j-1}$ to $n_{j}$ is
\begin{equation}\label{eq:step}
 T^\rho_{j}=\frac{1}{1-\rho_j}\matrise{1 & -\rho_j \\ -\rho_j & 1},
\end{equation}
where 
\begin{equation}\label{fresnel}
\rho_j(\omega)=\frac{n_{j-1}(\omega)-n_{j}(\omega)}{n_{j-1}(\omega)+n_{j}(\omega)}
\end{equation}
is the Fresnel reflection coefficient, associated with the index step between $n_{j-1}(\omega)$ and $n_{j}(\omega)$. The transfer matrix of the pure propagation in layer $j$ is
\begin{equation}\label{eq:prop}
 T^d_{j}=\matrise{\exp[i\omega n_j(\omega)d_j/c] & 0 \\ 0 & \exp[-i\omega n_j(\omega)d_j/c]},
\end{equation}
where $c$ is the vacuum light velocity. Note that Eqs. (\ref{eq:step})-(\ref{eq:prop}) are also valid when the refractive index is complex. The refractive index is necessarily complex for dispersive materials, because dispersion is accompanied by loss according to the Kramers--Kronig relations. The transfer matrix of the total structure is given by
\begin{equation}
M = T^d_{N}T^\rho_{N} T^d_{N-1}\cdots T^\rho_{2}T^d_1 T^\rho_1 T^d_0.
\end{equation}
The reflection coefficient $r(\omega)$ from the left, and the transmission coefficient $t(\omega)$ for the electric field from left to right, are given by the (2,1)- and (2,2)-elements of $M$:
\begin{subequations}
\begin{align}
 r(\omega) &= -\frac{M_{21}}{M_{22}}, \\
 t(\omega) &= \frac{\det M}{M_{22}}.
\end{align}
\end{subequations}

We will now describe a layer-peeling method that can be used in the presence of weak dispersion. The precise condition for the dispersion will become clear below. To be able to reconstruct the structure, we assume that the refractive index $n_0(\omega)$ of the zeroth layer and the reflection spectrum of the entire structure (as seen from $z=0$), are known. The goal is to calculate $n_j(\omega)$ for all $j \geq 1$ and $d_j$ for $j \geq 0$.

The reflection spectrum of the layered structure can be expressed as follows:
\begin{equation}
 r(\omega)=\int_{2d_0/c}^\infty h(t)\exp(i\omega t)\diff t,
\label{fourier}
\end{equation}
where $h(t)$ is the impulse response, i.e., the time-domain reflected field when a Dirac delta pulse is incident to the structure. Note the lower limit $2d_0/c$ in the integral: For a non-dispersive layer 0, the round-trip time to the first index step would be $2n_0d_0/c$; however, due to dispersion we can only be sure that the round-trip time is no less than $2d_0/c$.

We take the forward and backward propagating frequency-domain fields at $z=0$ to be $1$ and $r(\omega)$, respectively, defining the field vector 
\begin{equation}
\matrise{u_0(\omega) \\ v_0(\omega)}=\matrise{1 \\ r(\omega)}.
\end{equation}
For now we assume that the layer thicknesses $d_j$ are known {\it a priori}; the case with unknown layer thicknesses is treated below. The field vector before the beginning of the first layer ($z=d_0^-$) is given by 
\begin{equation}
\matrise{u_1(\omega) \\ v_1(\omega)} = T^d_0 \matrise{u_0(\omega) \\ v_0(\omega)}.
\end{equation}
We now define the reflection spectrum after the zeroth layer has been ``peeled off'':
\begin{equation}
r_1(\omega)=\frac{v_1(\omega)}{u_1(\omega)}.
\end{equation}

The index step at $z=d_0$ can be regarded as a frequency-dependent reflector with (unknown) reflectivity $\rho_1(\omega)$, in accordance with Eq. \eqref{fresnel}. We assume that the dispersions of layers 0 and 1 are sufficiently small, such that the time-domain response associated with $\rho_1(\omega)$ has duration less than $2d_1/c$. Then we can write 
\begin{equation}
 r_1(\omega)=\rho_1(\omega) + \int_{2d_1/c}^\infty h_1(t)\exp(i\omega t)\diff t,
\label{fourier1}
\end{equation}
where 
\begin{equation}
h_1(t)= \frac{1}{2\pi}\int_{-\infty}^\infty r_1(\omega)\exp(-i\omega t)\diff\omega.
\label{h1}
\end{equation}
In Eq. \eqref{fourier1} the lower limit in the integral reflects the fact that any reflections from the later index steps are delayed by (at least) the round-trip time $2d_1/c$.

Having established Eq. \eqref{fourier1}, we can now identify $\rho_1(\omega)$:
\begin{equation}
 \rho_1(\omega) = \int_{0}^{2d_1/c} h_1(t)\exp(i\omega t)\diff t.
\label{rho1principle}
\end{equation}
With the local reflection coefficient $\rho_1(\omega)$ in hand, we can calculate the refractive index of layer 1 using Eq. \eqref{fresnel}. Once $n_1(\omega)$ has been found, we can calculate the reflection spectrum with the first layer removed: 
\begin{equation}\label{eq:lp}
r_2(\omega)=\frac{v_2(\omega)}{u_2(\omega)}, \quad\matrise{u_2(\omega) \\ v_2(\omega)} = T^d_1 T^\rho_1 \matrise{u_1(\omega) \\ v_1(\omega)}.
\end{equation}
Now we find ourselves in the same situation as before, so we can continue the process until all layers have been found.

From Eq. \eqref{rho1principle} one obtains the \emph{complex} reflection coefficient of each layer, and therefore, by Eq. \eqref{fresnel}, both the real and imaginary parts of the refractive index. One may ask if the reconstructed refractive index automatically is causal, or whether the Kramers--Kronig relations could be used in addition to ensure causality. The answer is that the lower limit in the integral \eqref{rho1principle} ensures that $\rho_1(\omega)$ is causal, and therefore by Eq. \eqref{fresnel} that $n_1(\omega)$ is analytic in the upper half-plane of complex frequency. Provided $n_1(\omega)\to 1$ as $\omega\to\infty$, $n_1(\omega)$ is therefore guaranteed to be causal \cite{nussenzveig}.

In practical situations the available bandwidth is finite, the layer thicknesses may not be known \emph{a priori}, and the reflection data contains noise. We will now consider these aspects.

\subsection{Effect of finite bandwidth}
In practice, we only have reflection data in a limited bandwidth $\omega_1\leq\omega\leq\omega_2$. In other words, the reflection data $r_1(\omega)$ in Eq. \eqref{h1} is necessarily multiplied by a window function $W(\omega)$, which is nonzero only in the interval $\omega_1\leq\omega\leq\omega_2$. Physically, this means that instead of probing the structure with a Dirac delta pulse, we use an input pulse $w(t)$ of duration $\tau\sim2\pi/(\omega_2-\omega_1)$:
\begin{equation}
 w(t)=\frac{1}{2\pi}\int_{\omega_1}^{\omega_2} W(\omega)\exp(-i\omega t)\diff\omega.
\end{equation}
We require the duration of the time-domain response associated with $\rho_1(\omega)W(\omega)$ to be less than $2d_1/c$, in order to distinguish between the response due to the first and the other layers. The time-domain response associated with $\rho_1(\omega)$ may already have duration up to $\sim 2d_1/c$, so we must require $\tau \ll 2d_1/c$, or equivalently, $\omega_2-\omega_1 \gg \pi c/d_1$. This must be true for all layers, so
\begin{equation}\label{condmind}
\omega_2-\omega_1 \gg \frac{\pi c}{d_{\min}},
\end{equation}
where $d_{\min} \le \min_j d_j$ is a lower bound for the (possibly unknown) layer thicknesses. In addition to condition \eqref{condmind}, we recall that the time-domain response associated with $\rho_1(\omega)$ must have duration less than $2d_{\min}/c$, which means that the minimum allowable layer thickness is limited by the narrowest dispersion feature in the structure.

The response $h_1(t)*w(t)$ (where $*$ denotes convolution) is no longer guaranteed to vanish for $t<0$, so we must extend the lower integration limit in Eq. \eqref{rho1principle} to contain the pulse $w(t)$:
\begin{equation}
 \rho_1(\omega)W(\omega) = \int_{t_w}^{2d_1/c+t_w} h_1(t)*w(t)\exp(i\omega t)\diff t.
\label{rho1wprinciple}
\end{equation}
Here $t_w$ is the ``start position'' of $w(t)$, i.e., a possibly negative number such that the response from the first layer roughly is contained in the interval $[t_w, 2d_1/c+t_w]$. Eq. \eqref{rho1wprinciple} leads to the result $\rho_1(\omega)W(\omega)$ rather than $\rho_1(\omega)$; i.e., the response of the single layer has been filtered with $W(\omega)$. Thus we must make sure that the bandwidth $[\omega_1,\omega_2]$ of the window function matches that of the dispersion of $n_1(\omega)$ to be reconstructed.

\subsection{Unknown layer thicknesses}\label{sec:layer_thickness}
We will now describe how the layer peeling algorithm also can be used when the layer thicknesses are not known {\it a priori}. Recall that the time-domain responses associated with $\rho_j(\omega)$ have duration less than $2d_\text{min}/c$ for all $j$. Starting at a layer boundary, one can then perform layer peeling a distance $d_\text{min}$ using Eq. \eqref{eq:lp}, and calculate the resulting time-domain response. This removes the effect of the layer boundary, and the first signal in the transformed time-domain response is due to reflection from the next layer boundary. Let $t_i$ denote the start position of this signal and let $t_w$ denote the start position of $w(t)$. If $t_i>t_w$, this indicates that $d_\text{min}$ is less than the layer thickness. Define a small thickness $\Delta$, which is sufficiently small to achieve the desired, spatial resolution. Then we can transform the fields successively using Eq. \eqref{eq:prop} with the small thickness $\Delta$ until $t_i=t_w$, and we have arrived at an 
index step. We can then peel off the dispersive response associated with the index step and search for the next layer boundary, and so forth. If $W(\omega)$ only is nonzero in the interval $\omega_1\leq\omega\leq\omega_2$, the time-domain responses do not have well-defined fronts, and the procedure above for finding the layer thickness is ambiguous. However, one can use an alternative definition for the start position of the time domain signals, as shown in the numerical examples.

\subsection{Effect of noise}\label{sec:noise}
For any real system there is a given signal-to-noise ratio, which may be frequency dependent \cite{jepsen}. The layer peeling algorithm will fail if the reflection signal at a given index step becomes less than the noise. This can be due to either low Fresnel reflection from the index step itself, or high reflection or material absorption in the preceding part of the structure. The case with high reflection in the preceding part of the structure was analyzed in Ref. \cite{skaar_feced}, and it was shown that the noise amplification factor during the layer peeling algorithm was of the order of $1/T_{\min}$ where $T_{\min}$ is the minimum power transmission through the structure. A similar conclusion can be reached by considering the effect of absorption. Let $\varrho$ denote the minimum detectable reflection coefficient. The maximum probing depth, $d$, into a material with a single index step can be estimated by
\begin{equation}
\exp[-2\text{Im}(n)\omega d/c]\frac{\Delta n}{2n}\approx \varrho,
\end{equation}
where $\Delta n$ is the change in the real part of the refractive index and $n$ is the average real part of the refractive index at the step. Solving for the the maximum probing depth, we obtain
\begin{equation}\label{eq:noise}
d\approx\frac{c}{2\omega\text{Im}(n)}\ln\left(\frac{\Delta n}{2n\varrho}\right).
\end{equation}
We observe that the maximum probing depth into the structure is inversely proportional to the material absorption.
Assuming $\varrho=10^{-4}$, $\omega=2\pi\cdot 10^{12}\;\text{s}^{-1}$, $\Delta n/(2n)=10^{-2}$, and $\text{Im}(n)=0.01$ gives $d\approx1\;\text{cm}$, which corresponds to $\sim 30$ wavelengths. This is roughly the maximum depth one can expect the layer peeling algorithm to work for a material with comparatively low loss in the THz-range.

\section{Numerical examples}
\begin{figure}
\centerline{\includegraphics[width=10cm]{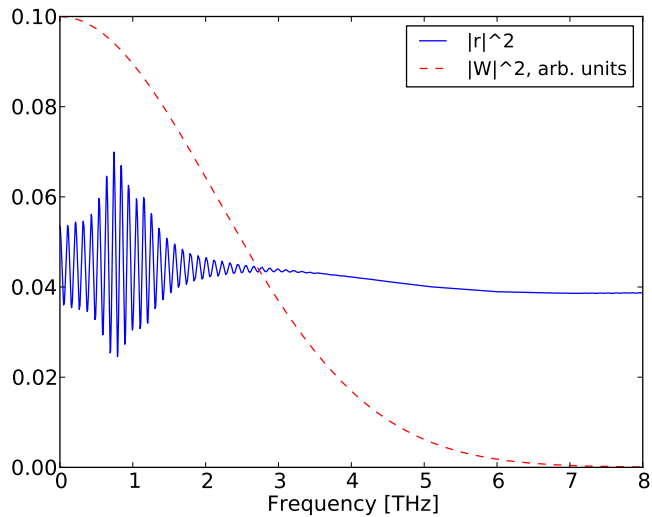}}
\caption{Power reflection coefficient and squared magnitude of the window function for the numerical example.} 
\label{fig:ref_w}
\end{figure}

\begin{figure}
\centerline{\includegraphics[width=10cm]{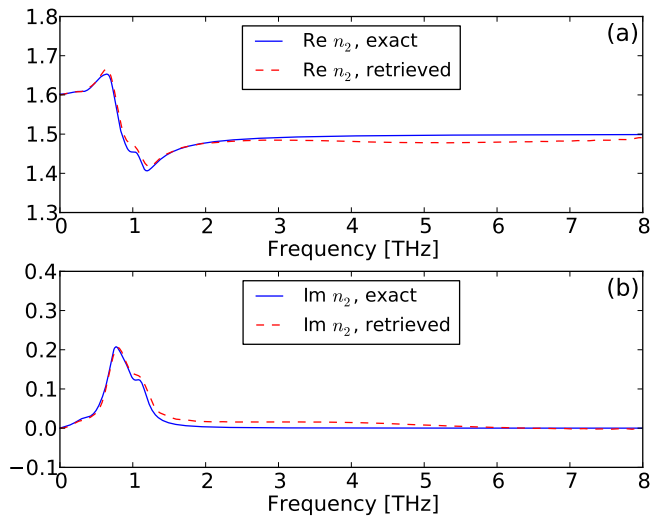}}
\caption{Exact and retrieved refractive index of the second layer. (a) Real part and (b) imaginary part of refractive index.} 
\label{fig:2nd_layer}
\end{figure}

As a first numerical example of the algorithm in Sec. \ref{layerpeelingdisp}, we consider a structure consisting of two material layers. We assume vacuum for $z<0$, a material with refractive index $n_1$ for $0\leq z<d_1$, and a material with
refractive index $n_2$ for $z\geq d_1$. Both materials are assumed to be dispersive and lossy. The task is to determine $n_1$, $n_2$, and $d_1$, given knowledge of the reflection coefficient at $z=0^-$ in a finite frequency range. The refractive index of the first material is assumed to be in the form
\begin{equation}\label{eq:n_mat}
n(\omega)=n_c\sqrt{1+\frac{\chi(\omega)}{n_c^2}},
\end{equation}
where $n_c$ is constant, and $\chi(\omega)$ represents a Lorentzian absorption feature, given by
\begin{equation}
\chi(\omega)=\frac{F\omega_0^2}{\omega_0^2-\omega^2-iG\omega}.
\end{equation}
We take $n_c=1.5$, $\omega_0=5\omega_s$, $F=0.1$, and $G=5\omega_s$ for the first material. 
Here $\omega_s=2\pi f_s$, where $f_s$ is a scaling frequency. Note that $n_c$ must approach $1$ as $\omega\rightarrow\infty$ for the refractive index in Eq. \eqref{eq:n_mat} to be causal. However, we here use the approximation that $n_c$ is constant in the frequency range $\omega_1\leq\omega\leq\omega_2$ and assume that $n_c$ has the correct behavior as $\omega\rightarrow\infty$. The refractive index of the second material is also assumed to be in the form Eq. (\ref{eq:n_mat}), with $n_c=1.5$, but here the susceptibility is taken to be the sum of ten Lorentzian absorption features of various amplitudes, bandwidths, and center frequencies, in the vicinity of $\omega_{0}=0.8\omega_s$. The resulting refractive index is seen in Fig. \ref{fig:2nd_layer}. In this example, we set $f_s=1$~THz, which leads to the vacuum wavelength $\lambda_s=c/f_s=0.3$ mm. We take $d_1=3\lambda_s$, and assume that the reflection coefficient is known in the frequency range 0--8~THz. The resulting power reflection coefficient is shown 
in Fig. \ref{fig:ref_w}.

In the layer peeling algorithm, we take $d_\text{min}=2\lambda_s$. In addition, we must choose an appropriate window function. The window function should have negligible energy outside the frequency window $[\omega_1,\omega_2]=[-8~\text{THz}, 8~\text{THz}]$. Additionally, the corresponding time-domain pulse should have a well-defined front. As a compromise between these two conflicting requirements, we use a Gaussian window function, defined in the time-domain by
\begin{equation}\label{eq:td_win}
w(t)=\cos(\omega_ct)\exp[-(t/\tau)^2].
\end{equation}
The width $\tau$ of the pulse and its central frequency $\omega_c$ are chosen to match its spectrum to the frequency range where the reflection coefficient is known. We use $\tau=0.08$ ps and $\omega_c=\omega_s$, which gives the window function $W(\omega)$ in Fig. \ref{fig:ref_w}. This window function has a small energy outside $[\omega_1,\omega_2]$. We here set $r(\omega)=r(\omega_2)$ for $|\omega|>\omega_2$ in the numerical implementation of the algorithm, in accordance with the assumption that the reflection spectrum outside $[\omega_1,\omega_2]$ is unknown. The lower integration limit in Eq. (\ref{rho1wprinciple}) is set to $t_w=-0.3$~ps.

\begin{figure}
\centerline{\includegraphics[width=10cm]{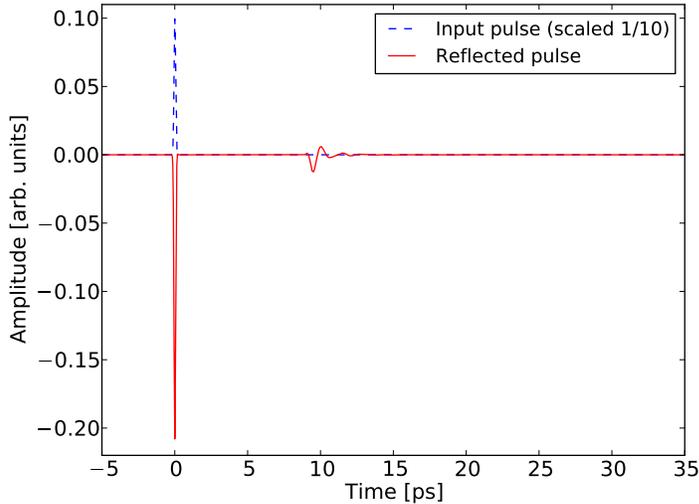}}
\caption{Incident and reflected pulse at $z=0^-$. The incident pulse is given by Eq. \eqref{eq:td_win}, which is the time-domain representation of the window function. The reflected pulse is given by the convolution of the incident pulse and $h_1(t)$. The amplitude of the incident pulse has been scaled a factor 1/10 in the figure.} 
\label{fig:pulses}
\end{figure}
\begin{figure}
\centerline{\includegraphics[width=10cm]{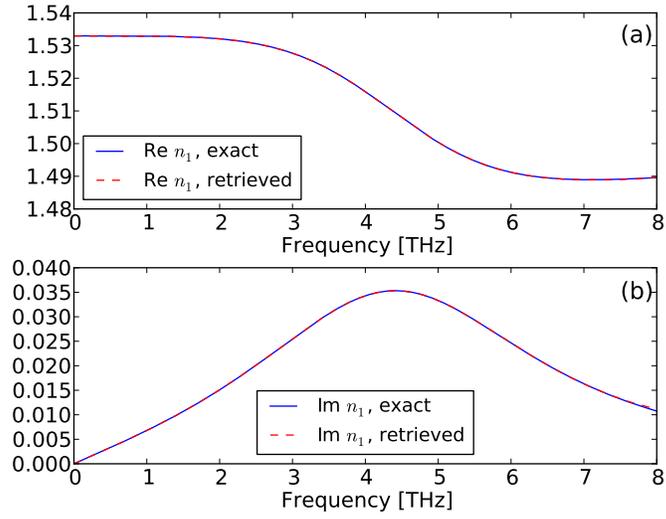}}
\caption{Exact and retrieved refractive index of the first layer. (a) Real part and (b) imaginary part of refractive index. The maximum error in the retrieved refractive index is $6\cdot 10^{-4}$.} 
\label{fig:1st_layer}
\end{figure}

Figure \ref{fig:pulses} shows $w(t)$, which represents the incident pulse used to probe the structure. Also shown is the convolution of the incident pulse and $h_1(t)$, which represents the reflected signal from the structure. The reflected signal consists of two pulses, where the first pulse is due to reflection at $z=0$ and the second pulse is due to reflection at $z=d_1$. We observe that the duration of the first reflected pulse is similar to the duration of the incident pulse, which is due to the comparatively low dispersion of the first layer. Because $d_\text{min}\gg \pi c/(\omega_2-\omega_1)$, there is no problem to correctly retrieve the refractive index of the first layer, as shown in Fig. \ref{fig:1st_layer}. 

Having found the refractive index of the first layer, the next task is to find its thickness, which is done according to the procedure in Sec. \ref{layerpeelingdisp}. However, care must be taken to define the start position of the time domain signals, because they do not have a well-defined front, as discussed in Sec. \ref{sec:layer_thickness}. In the numerical implementation of the algorithm we define the start of the pulse as the first peak in the amplitude, given the amplitude is larger than a certain noise limit. As noted in Sec. \ref{sec:noise}, if the amplitude of the second reflected pulse becomes too small, the layer peeling algorithm fails to find the thickness of the first layer. One can also show that the layer thickness $d_1$ should be determined to within an accuracy of $\Delta\ll\pi c/\omega_2$. An incorrect layer thickness leads to oscillations in the retrieved refractive index \cite{duv:1999}. For the present parameters, the retrieved layer thickness is $3.007\lambda_s$. 

Once $d_1$ is found we can "peel off'' the effect of the first layer using Eq. (\ref{eq:lp}), which means that we transform the reflection coefficient to the position $z=d_1^-$. 
\begin{figure}
\centerline{\includegraphics[width=10cm]{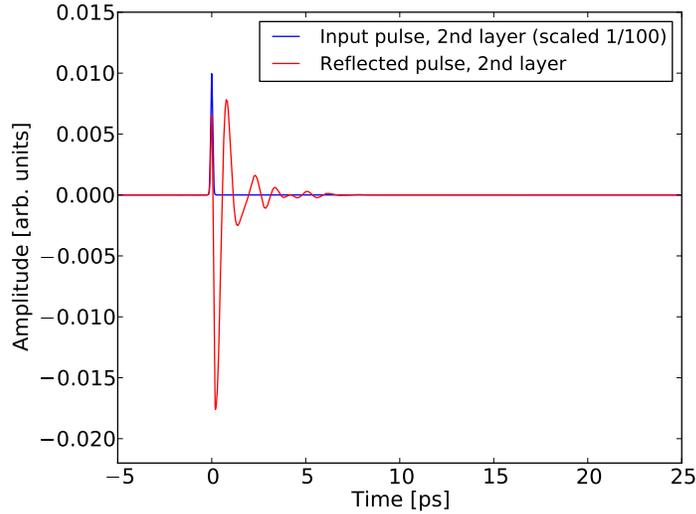}}
\caption{Incident and retrieved reflected pulse at $z=d_1^-$. The amplitude of the incident pulse has been scaled a factor 1/100 in the figure.} 
\label{fig:pulses2}
\end{figure}
Fig. \ref{fig:pulses2} shows the incident pulse, $w(t)$, and the reflected pulse from the second layer. The reflected pulse is calculated using the transformed reflection coefficient at the position $z=d_1^-$. We observe that the duration of the reflected pulse is significantly longer than that of the input pulse, which is due to the narrow dispersion feature of the second layer. The refractive index of the second layer is determined in the same manner as for the first layer, and the retrieved refractive index is shown in Fig. \ref{fig:2nd_layer}. The error in the retrieved refractive index of the second layer is mainly due to the small inaccuracy in the retrieved $d_1$.
\subsection{Effect of too small assumed layer thickness}
\begin{figure}
\centerline{\includegraphics[width=10cm]{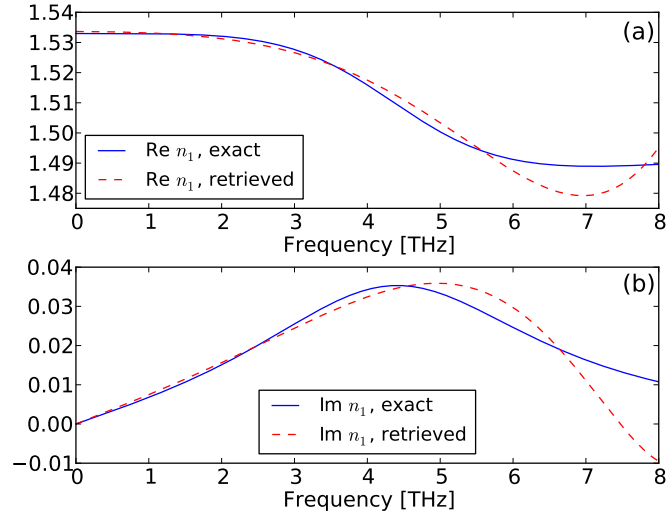}}
\caption{Exact and retrieved refractive index of the first layer when $d_\text{min}=0.25\lambda_s$. (a) Real part and (b) imaginary part of refractive index.} 
\label{fig:1st_layer_example1}
\end{figure}
As noted in Sec. \ref{layerpeelingdisp}, the duration of the time-domain response associated with reflection at an index step must be much less than $2d_\text{min}/c$ for the layer peeling algorithm to work. If material dispersion is weak, the duration of the time-domain response is mainly given by the duration of the window function $w(t)$. Assuming $\tau=0.08$ ps, as in the first example, we must chose $d_\text{min}\gg c\tau/2=0.04\lambda_s$ for successful retrieval of the refractive index in the first layer. As an example where $d_\text{min}$ is marginally too small, we set $d_\text{min}=0.25\lambda_s$, with otherwise the same parameters as in the first example. The retrieved refractive index of the first layer is shown in Fig. \ref{fig:1st_layer_example1}. We observe that the retrieved refractive index deviates significantly from the exact refractive index. In addition, the layer peeling algorithm fails to find the correct layer thickness in this case due to the residual time-domain response in $h_1(t)$ 
caused by incomplete removal of the first layer.

\subsection{Effect of noise}
\begin{figure}
\centerline{\includegraphics[width=10cm]{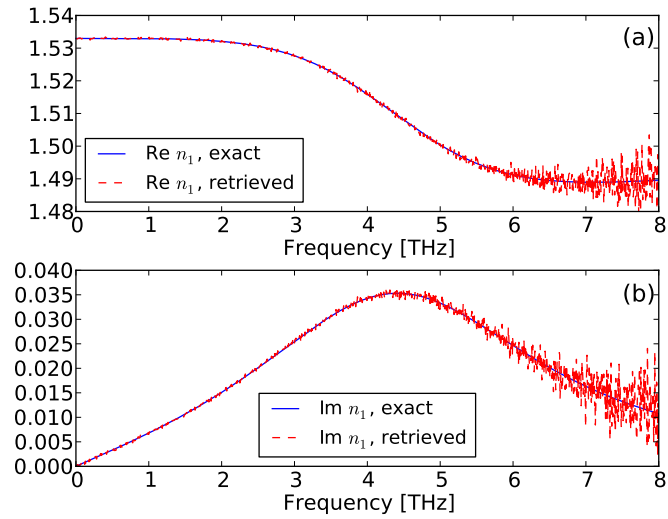}}
\caption{Exact and retrieved refractive index of the first layer in the presence of noise. (a) Real part and (b) imaginary part of refractive index.} 
\label{fig:1st_layer_example2}
\end{figure}
\begin{figure}
\centerline{\includegraphics[width=10cm]{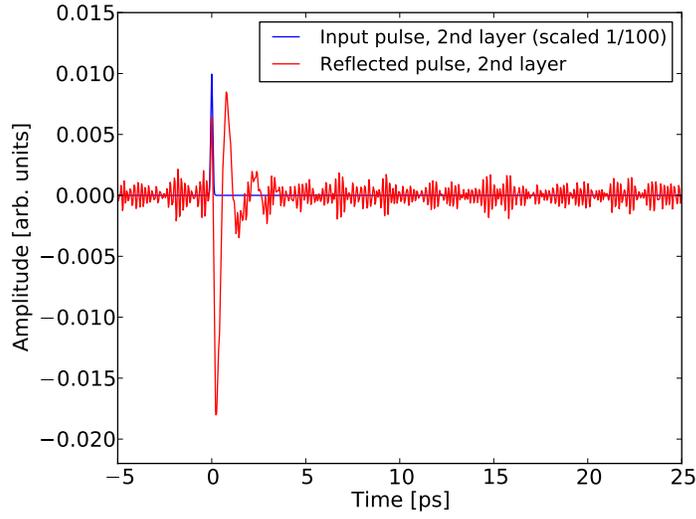}}
\caption{Incident and retrieved reflected pulse at $z=d_1^-$ in the presence of noise. The incident pulse (including the noise) has been scaled a factor 1/100 in the figure.} 
\label{fig:pulses2_example_2}
\end{figure}
\begin{figure}
\centerline{\includegraphics[width=10cm]{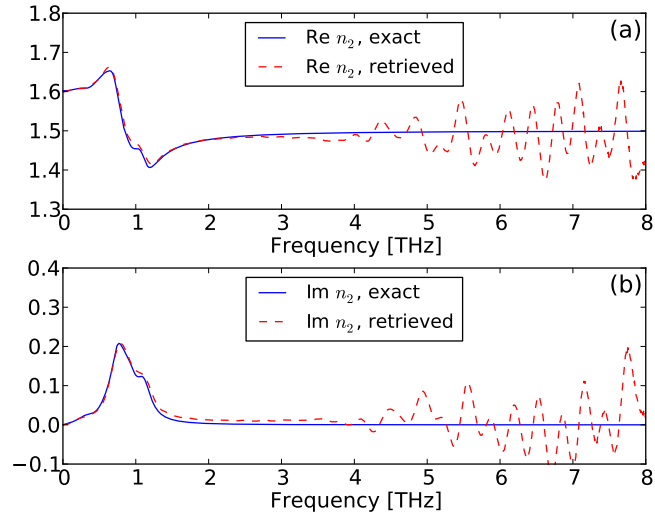}}
\caption{Exact and retrieved refractive index of the second layer in the presence of noise. (a) Real part and (b) imaginary part of refractive index.} 
\label{fig:2nd_layer_example_2}
\end{figure}
As an example of the influence of noise on the layer peeling algorithm, we consider the same structure as in the first example, but with noise added to the input pulse $w(t)$. The noise is assumed to be white and Gaussian, with mean value 0, standard deviation $5\cdot 10^{-5}$, and time $1/(200~\text{THz})$ between the samples. The signal to noise ratio is thus proportional to $W(\omega)$ in the frequency domain. The retrieved refractive index of the first layer is shown in Fig. \ref{fig:1st_layer_example2}. As expected, the influence of the noise is most severe at high frequencies, where the signal-to noise ratio is small. We can estimate the maximum probing depth using Eq. (\ref{eq:noise}). Because the loss is not constant over the frequency range where the reflection coefficient is known, we use the approximate values at $f= 4$ THz, where $\text{Im}(n)\approx 0.035$, $\Delta n/(2n)\approx 0.01$, and $\varrho\approx 5\cdot 10^{-5}$, giving $d\approx 3\lambda_s$ for the maximum layer thickness of the first 
layer where we can 
expect the layer peeling algorithm to work. Figure $\ref{fig:pulses2_example_2}$ shows the reflected pulse from the second layer, as determined from the layer peeling algorithm, and Fig. $\ref{fig:2nd_layer_example_2}$ shows the retrieved refractive index of the second layer, in the presence of noise. We observe that the retrieved refractive index is erroneous above 4 THz, which is due to the low signal to noise ratio at increasing frequency. When the thickness of the first layer becomes much larger than 3$\lambda_s$, the reconstructed refractive index also becomes inaccurate at lower frequencies.

\subsection{Structure with three layers}
\begin{figure}
\centerline{\includegraphics[width=10cm]{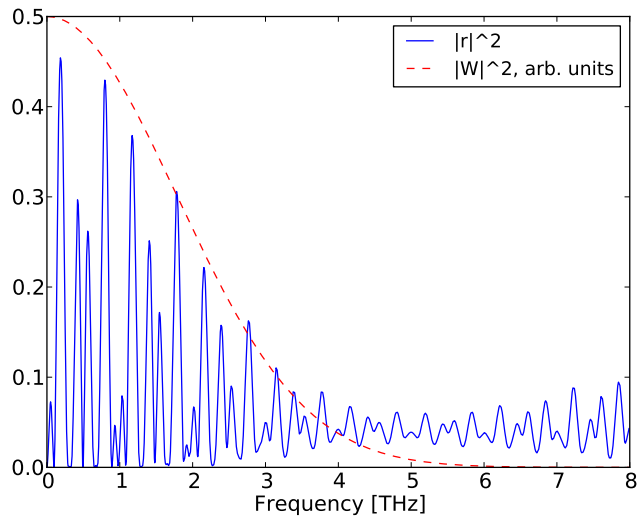}}
\caption{Power reflection coefficient and squared magnitude of window function for the structure with three layers.} 
\label{fig:ref_w_example_3}
\end{figure}
\begin{figure}
\centerline{\includegraphics[width=10cm]{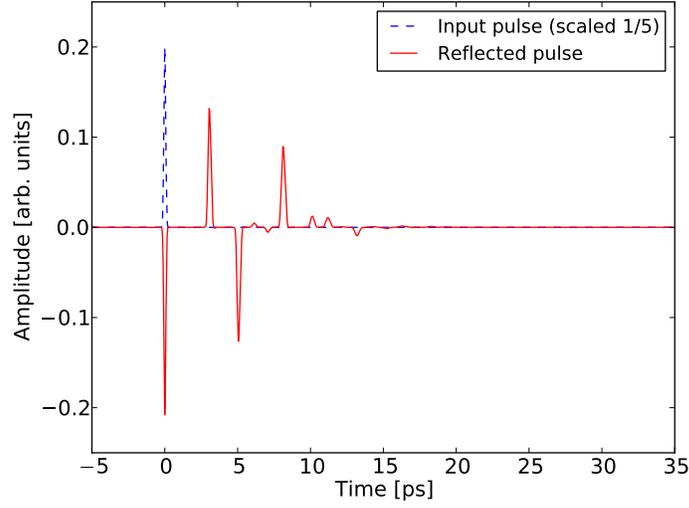}}
\caption{Incident and reflected pulse at $z=0^-$ for the structure with three layers. The amplitude of the incident pulse has been scaled a factor 1/5 in the figure.}
\label{fig:pulses_example_3}
\end{figure}
\begin{figure}
\centerline{\includegraphics[width=10cm]{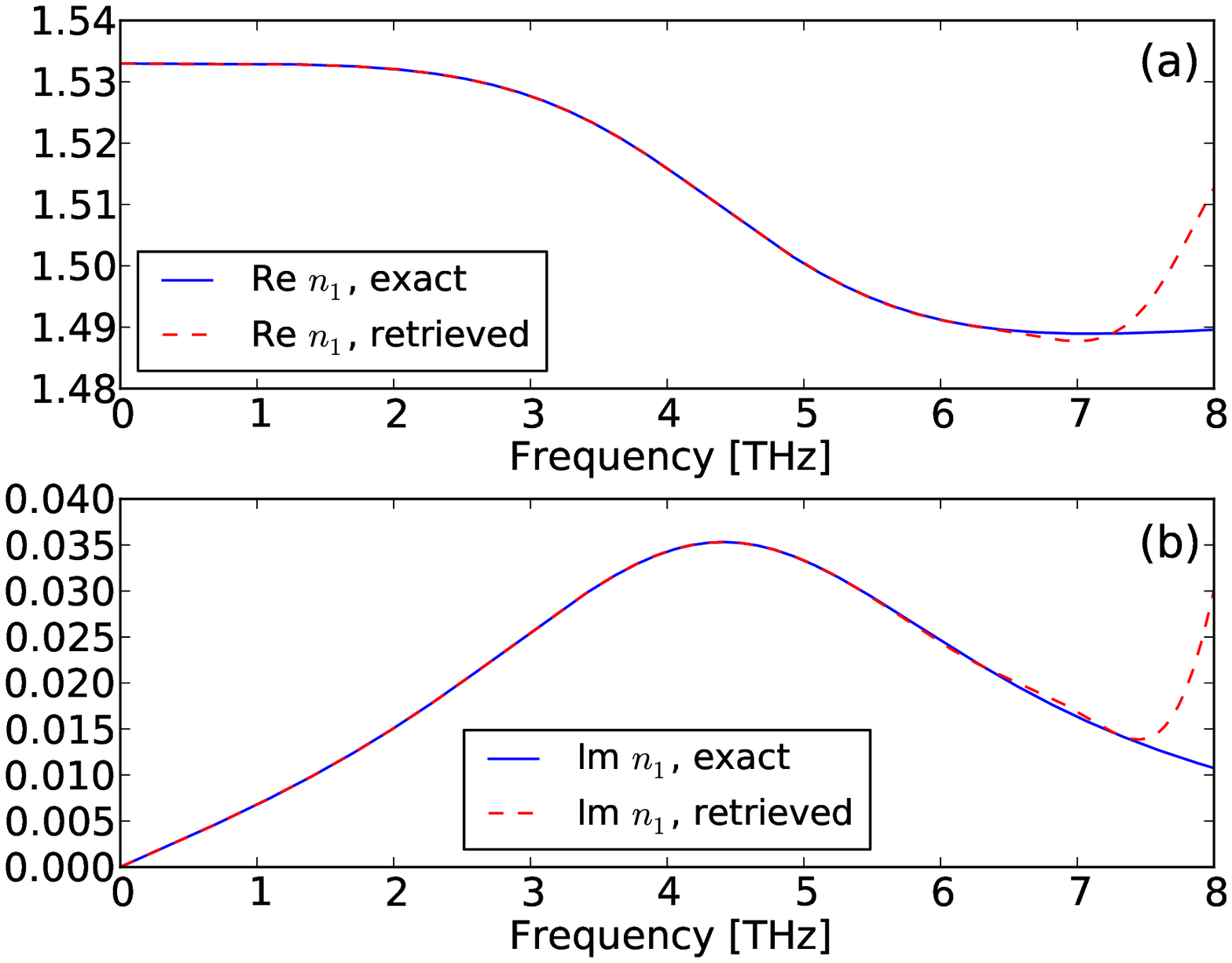}}
\caption{Exact and retrieved refractive index of the first layer for the structure with three layers. (a) Real part and (b) imaginary part of refractive index.} 
\label{fig:1st_layer_example3}
\end{figure}
\begin{figure}
\centerline{\includegraphics[width=10cm]{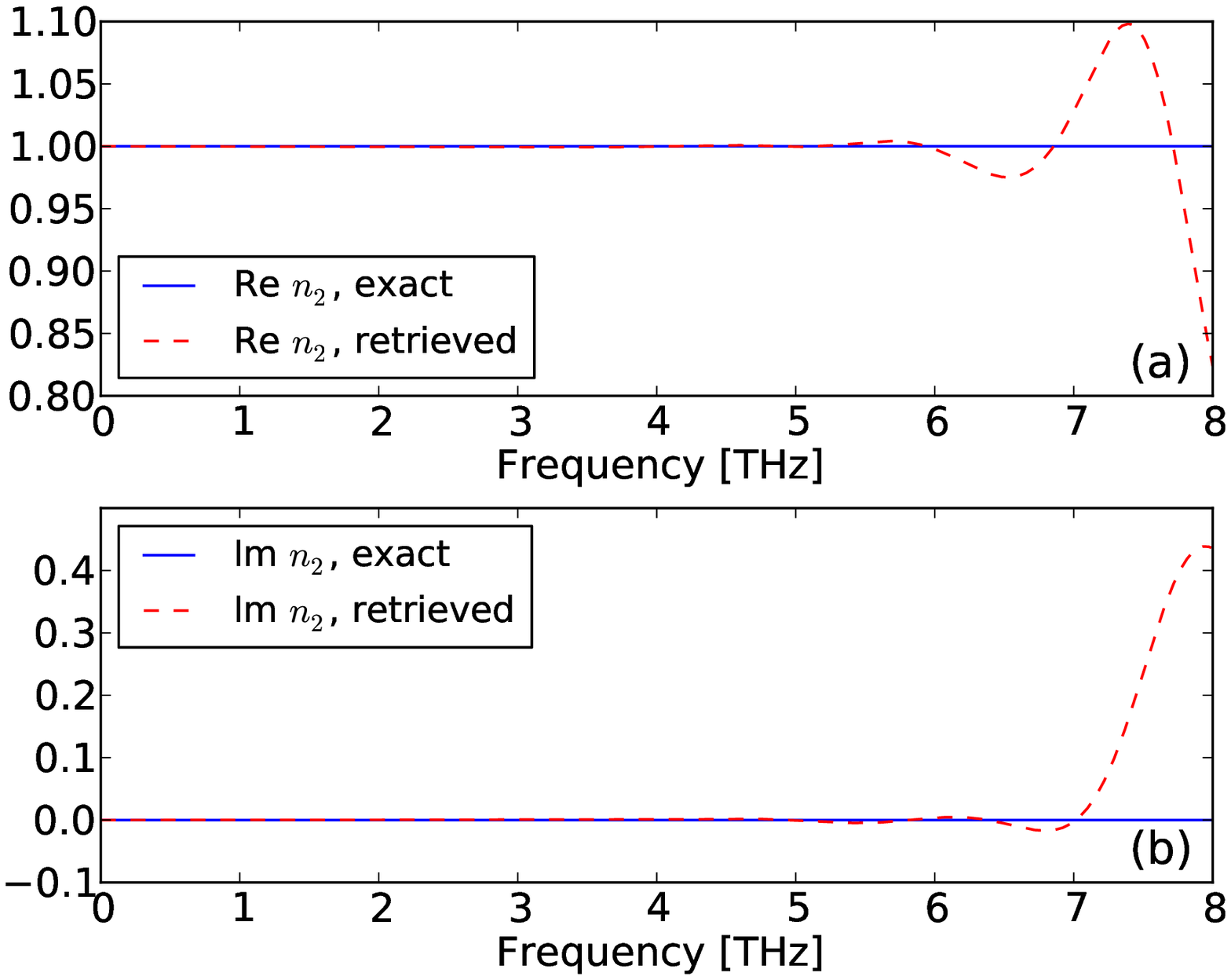}}
\caption{Exact and retrieved refractive index of the second layer for the structure with three layers. (a) Real part and (b) imaginary part of refractive index.} 
\label{fig:2nd_layer_example3}
\end{figure}
\begin{figure}
\centerline{\includegraphics[width=10cm]{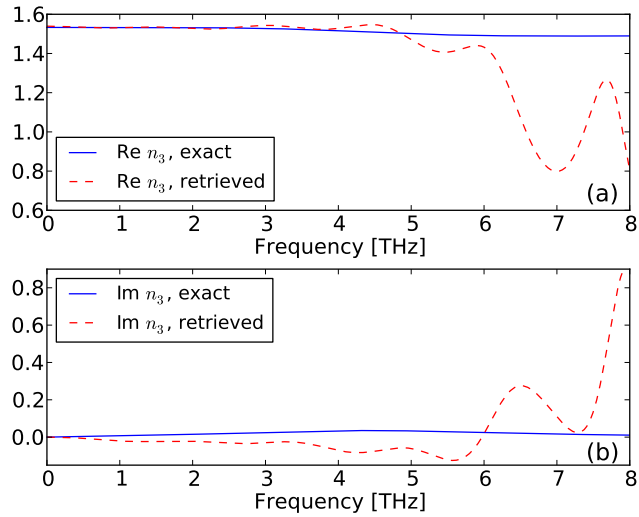}}
\caption{Exact and retrieved refractive index of the third layer. (a) Real part and (b) imaginary part of refractive index.} 
\label{fig:3rd_layer_example3}
\end{figure}
As a final example, we will consider a structure with three layers, with vacuum for $z<0$. The refractive index of the first and third layer is the same as in the first example, and the refractive index of the second layer is 1. The thickness of the first and second layer is $d_1=d_2=\lambda_s$, with $d_3=\infty$. We take $d_\text{min}=\lambda_s/2$ in the layer peeling algorithm and use the same window function as in the first example. Figure \ref{fig:ref_w_example_3} shows the power reflection coefficient of the structure. The corresponding time-domain pulses reflected from the structure are shown in Fig. \ref{fig:pulses_example_3}. Using the layer peeling algorithm we retrieve the refractive indices and layer thicknesses of the structure. The retrieved refractive index of layer 1--3 are shown in Figs. \ref{fig:1st_layer_example3}-\ref{fig:3rd_layer_example3}, respectively. We observe that there are errors in the retrieved refractive index at high frequencies. The errors increase and approach shorter 
frequencies for the successive layers. This is due to the unknown reflection coefficient above 8 THz. Because the window function has a small energy above this frequency, the unknown part of the reflection coefficient causes an unphysical precursor to each reflected pulse in the time-domain response. The errors in retrieved refractive index are here due to the overlap between the main pulse from one layer and the precursor from the pulse reflected from the next layer. For the present parameters, this type of error in the retrieved refractive index is strongly reduced if the reflection coefficient is assumed known up to 12 THz instead of 8 THz. We also find that an incorrect retrieved thickness of the first layer will lead to increasingly large errors in the retrieved refractive indices of the following layers. For the present parameters, an error of $10^{-3}\lambda_s$ in $d_1$ leads to large errors in the retrieved $n_2$ and especially $n_3$.

\section{Impossibility of inverse scattering for large dispersion}
We will now prove that for dispersive structures, the inverse scattering problem does not have a unique solution in general. To this end, we consider the Fresnel reflection coefficient $\rho_1(\omega)$ associated with a \emph{single} index step, from $n_0(\omega)$ to $n_1(\omega)$, and prove that this reflection coefficient can also be realized as a lossless and dispersionless structure; a stack of discrete reflectors in vacuum, or equivalently, a layered structure. 

Let $\bar\rho_1^1$, $\bar\rho_1^2$, $\ldots$ be the reflection coefficients of the discrete reflectors, and $\Delta$ the distance between the reflectors. We choose
\begin{equation}
\frac{2\Delta}{c}=\frac{2\pi}{2\omega_{\max}},
\end{equation}
in accordance with the considered bandwidth, from $-\omega_{\max}$ to $\omega_{\max}$. Initially we set $\rho_1^1(\omega)=\rho_1(\omega)$. By causality
\begin{equation}
 \rho_1^1(\omega) = \int_{0}^{\infty} h_1^1(t)\exp(i\omega t)\diff t
\label{rho1principle2}
\end{equation}
for a real time-domain response $h_1^1(t)$. 

We now discretize the continuous response $h_1^1(t)$, leading to $h_1^1[j]$:
\begin{equation}
 \rho_1^1(\omega) = \sum_{j=0}^{\infty} h_1^1[j]\exp(i\omega j2\Delta/c)
\label{rho1principle3}
\end{equation}
in the bandwidth $|\omega|\leq \omega_{\max}$. If $\rho_1^1(\omega)$ were zero outside this bandwidth, the Nyquist sampling theorem would immediately give the connection between $h_1^1[j]$ and $h_1^1(j 2\Delta/c)$. In general, however, the exact relation is found by extending $\rho_1^1(\omega)$ to a periodic function with period $2\omega_{\max}$, and setting $h_1^1[j]$ equal to the associated Fourier coefficients. With this procedure, Eqs. \eqref{rho1principle2} and \eqref{rho1principle3} would be identical in the relevant bandwidth, however with lower limit $-\infty$ in the sum \eqref{rho1principle3}. Setting the lower limit to 0 amounts to finding the optimal causal and discrete approximation to expression \eqref{rho1principle2}. In the limit where $\rho_1^1(\omega)$ vanishes for $|\omega|>\omega_{\max}$, the error in the approximation tends to zero. 

Assuming that the discrete response $h_1^1[j]$ is the reflection from a stack of discrete, lossless reflectors, we can perform layer peeling identifying the reflectors. By causality the first reflector is given by
\begin{equation}
 \bar\rho_1^1 = h_1^1[0].
\label{firstrefl}
\end{equation}
Note that the reflector $\bar\rho_1^1$ is real, since the time-domain response $h_1^1(t)$, and therefore $h_1^1[j]$, is real: $h_1^1(t)$ is a physical time-domain response as resulting from a real impulse. Peeling off this reflector, and removing the subsequent pure propagation in the layer with thickness $\Delta$, can be done with the associated transfer matrices, or equivalently, by applying the Schur recursion formula \cite{brucksteintrans,yagle_levy,skaar_erdogan}
\begin{equation}
 \rho_1^2(\omega) = \exp(-i\omega 2\Delta/c)\frac{\rho_1^1(\omega)-\bar\rho_1^1}{1-\bar\rho_1^1\rho_1^1(\omega)}.
\label{schurrecursion}
\end{equation}
The layer peeling process can be continued until all reflectors have been found. 

In other words, two different structures give the same reflection response $\rho_1(\omega)$; an index step between $n_0(\omega)$ and $n_1(\omega)$, and several layers with non-dispersive, real refractive indices. To be able to solve the inverse scattering problem of a dispersive structure, it is therefore apparent that extra information (in addition to the reflection spectrum) must be known. In Sec. \ref{layerpeelingdisp} we used the extra information that the dispersion is sufficiently small, such that the time-domain response associated with each Fresnel reflection has duration less than the round-trip time to the next index step. In addition we assumed $d_j \gg {\pi c}/{(\omega_2-\omega_1)}$ for all $j$.

\section{Conclusion}
An inverse scattering algorithm is applied to retrieve the material parameters of stratified structures. Even though this problem does not have a unique solution in general, there exist cases where the algorithm can be applied, given additional information about the structure. Specifically, for a given, lower bound of the layer thicknesses, the dispersion must be sufficiently small, and the frequency range where the reflection coefficient is known must be sufficiently large. The retrieval of material parameters of hidden layers is challenging due to absorption, noise, and unknown layer thicknesses. Despite these challenges, there exist cases where the algorithm is successful, as illustrated by the numerical examples.


\end{document}